\documentstyle[12pt]{article}

\newcommand{\EQ}{\begin{equation}}
\newcommand{\EN}{\end{equation}}

\newcommand{\bear}{\begin{eqnarray}}
\newcommand{\ear}{\end{eqnarray}}

\begin{document}

\topmargin 0pt
\oddsidemargin 5mm
\newcommand{\NP}[1]{Nucl.\ Phys.\ {\bf #1}}
\newcommand{\PL}[1]{Phys.\ Lett.\ {\bf #1}}
\newcommand{\NC}[1]{Nuovo Cimento {\bf #1}}
\newcommand{\CMP}[1]{Comm.\ Math.\ Phys.\ {\bf #1}}
\newcommand{\PR}[1]{Phys.\ Rev.\ {\bf #1}}
\newcommand{\PRL}[1]{Phys.\ Rev.\ Lett.\ {\bf #1}}
\newcommand{\MPL}[1]{Mod.\ Phys.\ Lett.\ {\bf #1}}
\newcommand{\JETP}[1]{Sov.\ Phys.\ JETP {\bf #1}}
\newcommand{\TMP}[1]{Teor.\ Mat.\ Fiz.\ {\bf #1}}

\renewcommand{\thefootnote}{\fnsymbol{footnote}}

\newpage
\setcounter{page}{0}
\begin{titlepage}
\begin{flushright}
UFSCARF-TH-94-11
\end{flushright}
\vspace{0.5cm}
\begin{center}
{\large  A note on graded Yang-Baxter solutions as braid-monoid invariants}\\
\vspace{1cm}
\vspace{1cm}
{\large M.J.  Martins  and P.B. Ramos} \\
\vspace{1cm}
{\em Universidade Federal de S\~ao Carlos\\
Departamento de F\'isica \\
C.P. 676, 13560~~S\~ao Carlos, Brasil}\\
\end{center}
\vspace{1.2cm}

\begin{abstract}
We construct two $Osp(n|2m)$ solutions of the graded Yang-Baxter equation by
using
the algebraic braid-monoid approach. The factorizable S-matrix interpretation
of these
solutions is also discussed.
\end{abstract}
\vspace{.2cm}
\vspace{.2cm}
\centerline{July 1994}
\end{titlepage}

\renewcommand{\thefootnote}{\arabic{footnote}}
\setcounter{footnote}{0}

\newpage
It is known that the Yang-Baxter equations play a central role in the study of
two-dimension exactly solvable models \cite{YB,KS,YB1}. One possible
generalization
of the Yang-Baxter relation is to consider integrable systems containing
both bosonic and fermionic degrees of freedom. In this case the elementary
generator
$A_{i}$ acts on a $Z_2$ graded vector space and its bosonic and fermionic
components
are distinguished by the parity $p(A_{i}) \equiv p(i) =0,1$, respectively
\cite{KS}.
Considering a graded space $V^{(n|m)}$ consisted of $n$ bosons and $m$
fermions, the graded
Yang-Baxter equation for the $R$-matrix amplitude is written as \cite {KS}
\EQ
R_{a_1,a_2}^{\alpha,\gamma}(u) R_{\alpha,a_3}^{b_1,\delta}(v+u)
R_{\gamma,\delta}^{b_2,b_3}(v)
(-1)^{p(\gamma)[p(a_3)+p(\delta)]} =
R_{a_2,a_3}^{\gamma,\delta}(v) R_{a_1,\delta}^{\alpha,b_3}(v+u)
R_{\alpha,\gamma}^{b_1,b_2}(u)
(-1)^{p(\gamma)[p(b_3)+p(\delta)]} \nonumber \\
\EN
where $p(i)=0$ for $i=1,2, \cdots, n$ ; $p(i)=1$ for $i=n+1,n+2, \cdots, n+m$.
It has also been
assumed that
the non-null elements $R_{a,b}^{c,d}$ are commuting variables, namely
$p(R_{a,b}^{c,d})=0$ \cite{KS}.
The simplest solution of the graded Yang-Baxter equation (1) was exhibited by
Kulish and Sklyanin \cite{KS} as the generalization of Yang's $S$-matrix
\cite{YY}, and is given by
\EQ
R(u,\eta) = \frac{u}{u+\eta} I + \frac{\eta}{u+\eta}P^{g}
\EN
where $I$ is the identity and $P^{g}$ is the graded permutation operator on the
tensor vector space
$V^{(n|m)} \otimes V^{(n|m)}$ with elements $(P^{g})_{a,b}^{c,d} =
(-1)^{p(a)p(b)} \delta_{a,d} \delta_{b,c}$.
.Variable $u$ is the spectral parameter while $\eta$ is a constant connected to
the graded classical
solution \cite{KS}.

In general, solutions of the $graded$ Yang-Baxter relation have been
investigated as invariants under the
superalgebras $Sl(n|m)$ and $Osp(n|2m)$ \cite{BS,KR,SA}. For instance,
trigonometric solutions have been constructed by
Bazhanov and Shadrikov \cite{BS} by
investigating the classical analog of the graded Yang-Baxter equations. The
special
case of the universal $Osp(2|1)$  $R$-matrix was discussed in ref. \cite{KR,SA}
in the context of the
quantum supergroup. Nowadays, however, it has been recognized that the
Yang-Baxter solutions are deeply
connected to a number of other algebraic structures, e.g.,
the braid-monoid \cite{JA}, the Temperely-Lieb(TL) \cite{TL}
algebras and more recently
the multi-colour versions of these structures \cite{MO}. In this sense one
would
expect that similar relations shall
also appear for the $graded$ case. In fact, Deguchi
and Akutsu \cite{DA} have shown that the fundamental
$Sl(n|m)$ graded solution can be obtained through the generators of the Hecke
algebra. Motived by this fact, the purpose
of this note is to discuss two $Osp(n|2m)$ solutions generated
by the braid-monoid invariants. We also comment on the crossing symmetry
property which is
fundamental in the context
of factorizable $S$-matrices interpretation of our solutions.

We start our discussion by constructing a $Osp(n|2m)$ TL invariant operator. In
order to build up
such operator
we recall that an $Osp(n|2m)$ invariant $A$ is a $(n+2m)X(n+2m)$
matrix satisfying the property (see e.g.
ref. \cite{CO})
\EQ
A +\alpha A^{st} \alpha^{-1} =0
\EN
where the symbol $A^{st}$ denotes the supertranspose
operation on the matrix $A$ and the matrix $\alpha$
is given by
\EQ
\alpha=\left( \begin{array}{cc}
	I_{nXn} &   O_{nX2m} \\
	O_{2mXn} & \left( \begin{array}{cc} O_{mXm} & I_{mXm} \\
	-I_{mXm} & O_{mXm} \\ \end{array} \right)\\
	\end{array}
	\right)
\EN
where $I_{aXa}$($O_{aXa}$) is the $aXa$
identity(null) matrix. Remarkably enough, we notice that the matrix
$\alpha$ present in $Osp(n|2m)$ invariance
plays a fundamental role on the construction of our TL
invariant. Indeed, if we define the following generator $E_i$ as
\EQ
E_i =  \sum_{abcd} \alpha_{ab} \alpha_{cd}^{st} e_{ac}^i \otimes e_{bd}^{i+1}
\EN
one can check that the Temperely-Lieb relations are satisfied, namely
\begin{eqnarray}
E_i E_{i \pm 1} E_i &= &E_i \nonumber \\
E_i^2 = (n-2m)E_i; & & [E_i,E_j]=0~~ for~~ |i-j| \geq 2
\end{eqnarray}
where the matrix elements of $e_{ab}^i$ acting on
$i^{th}$ ``site'' are $(e_{ab}^i)_{cd} = \delta_{a,c} \delta_{b,d}$.

The next step is to show how one can $graded$ ``Baxterize'' the explicit
representation (5) for the
monoid $E_i$. However, from the discussions of ref. \cite{KS,DA} we notice that
a null-parity graded
$R$-matrix satisfying (1) can be obtained by the relation
\EQ
R_i(u)= P_i^g X_i(u)
\EN
where $X_i(u)$ is a $null$-$parity$ usual Yang-Baxter operator satisfying the
relation
\EQ
X_i(u)X_{i+1}(u+v)X_i(v) =  X_{i+1}(v)X_i(u+v) X_{i+1}(u)
\EN

Finally, taking into account the previous experience
\cite{BA,PE} in the Baxterization of a TL generator
we find the following solution
\EQ
R_i(u,\eta) = P_i^g +f(u,\eta)E_i
\EN
where we have used the fact that $p(E_i)=0$ and the important identity
\EQ
P_i^g E_i =E_i P_i^g = E_i
\EN
and function $f(u,\eta)$ is given by
\EQ
f(u,\eta)= \left\{ \begin{array}{ccc}
	       \pm \frac{\sinh(u/\eta)}{\sinh(\gamma-u/\eta)} & \mbox{if
$2\cosh(\gamma)=(n-2m) >_< \pm 2$} \\
              \pm \frac{u}{\eta-u} & \mbox{ if $n-2m= \pm 2$} \\
                   \frac{\sin(u/\eta)}{\sin(\gamma-u/\eta)} & \mbox{ if
$2\cos(\gamma)= |n-2m|<2$}
\end{array}
\right.
\EN

We would like to stress that one advantage of this approach is that we are able
to generate
a new trigonometric/rational $Osp(n|2m)$ solution which
has no $graded$ classical analog, and apparently for
this
reason has been missed in the
literature \cite{BS}. From the point of view of quantum spin chains the
operator
$E_i$ generalizes previous effort in finding isotropic high-spins \cite{AF,BAT}
TL invariants. The fact that the
TL parameter $(n-2m)$ may assume negative values means that an appropriate
deformation of isotropic
high-spin chains \cite{BAT1,AL} shall possess indeed a hidden $Osp(n|2m)$
symmetry. For instance, we have
checked that the simplest case of $Osp(1|2)$ model corresponds to the deformed
point $q=i$ ( in the
notation of ref. \cite{BAT1}) of the spin-1 TL chain.

A second important feature of this approach is as follows.
First of all, identity (10) strongly suggests that the operators $P_i^g$ and
$E_i$ may be generators
of a more general algebraic structure, namely the braid-monoid algebra .
Moreover, taking into account
the remarks of ref. \cite{JA}, one can verify that a crossing symmetric
$S$-matrix interpretation of
(9) will lead us in the high energy
limit to the braid operator $P_i^g$ and
at the crossing point $u=\eta \gamma$ to the
monoid operator $E_i$. More precisely, one can show that besides equation (10)
we have the following
extra relations
\begin{eqnarray}
P_{i \pm 1}^g P_i^g E_{i \pm 1} &=& E_i P_{i \pm 1}^g P_i^g =  E_i E_{i \pm 1}
\\
E_i P_{i \pm 1} E_i & =& E_i
\end{eqnarray}
and the braid-inverse properties
\begin{eqnarray}
P_{i + 1}^g P_{i}^g P_{i + 1}^g & =& P_i^g P_{i+}^g P_i^g \\
 P_i^g P_i^g & = & I_i
\end{eqnarray}

In fact we can show that these set of relations between
the operators $P_i^g$ and $E_i$\footnote{At this point we recall the reader
that more general forms
of such monoid can be chosen. For instance, we mention  the monoid
$E_i=\sum_{abcd}
\alpha_{ab} \alpha_{cd}^{-1} e_{ac}^i \otimes e_{cd}^{i+1}$ where $\alpha
=diag[A_{nXn},antdiag(
B_{mXm},-B_{mXm})]$ if $A$ and $B$ are symmetric and invertible matrices.}
form a degenerated representation of a reduced\footnote{This occurs at the
singular point of the
parameters entering in the Birman-Wenzel algebra such that the eigenvalues
become degenerated.}
Birman-Wenzel algebra (see. e.g. \cite{JA,CH}). It is also possible to show
that the other relations
between the operators $P_i^g$ and $E_i$ closing the reduced Birman-Wenzel
algebra are just a
consequence of the identities (12,13,15). Hence, this
observation suggests that another $graded$ Baxterization can be implemented in
the sense of that
found by Jones \cite{JO}. Therefore,
proceeding as in the TL case and taking as a guess the Jones \cite{JO}
parametrization of the degenerated point of the Birman-Wenzel algebra \cite{CH}
we find that
\EQ
R_i(u,\eta)= \frac{u}{\eta}I_i +P_i^g -\frac{u}{u+\eta(n-2m-2)/2} E_i
\EN
satisfies the $graded$ Yang Baxter equation (1). A simple way to verify this
last result is
by checking that $P_i^g R_i(u,\eta)$ satisfies
the usual Yang-Baxter equation (8) if one uses the
relations (6,10,12-15). We recall, however,
that this solution corresponds to the rational limit
of a trigonometric $Osp(n|2m)$ solution already
found by Bazhanov and Shadrikov \cite{BS}. This is due to fact that (16) admits
its $graded$ classical
analog around the point $1/\eta \simeq 0$.

To conclude we would like to make some remarks concerning the interpretation of
the
graded solutions (9,16) as factorizable $S$-matrices. In order to interpret
$R_i(u,\eta)$ as a
$S$-matrix one has to impose crossing and unitarity conditions. Although the
unitarity condition remains
as usual, the crossing property in the graded case
has now to take into account the signs coming from the interchange of two
fermions. This is accomplished
by taking the supertranspose instead the traditional transpose operation, and
the crossing
symmetry property becomes
\EQ
S_i(\theta) =C \otimes I S_i^{st_i}(i\pi-\theta) (C \otimes I)^{st_i}
\EN
where $\theta$ is the relativistic rapidity and the supertranspose is taken
only on the first
space of $S_i(\theta)$. $C$ is the charge matrix, which for the theories (9,16)
is $C=\alpha$.
After some calculations, the corresponding $S$-matrix associated to the
solution (9) is
given by
\begin{eqnarray}
S_i(\theta) &= & f(\theta) \sin(\frac{\pi-i\theta}{\eta})
R(i\theta, \eta),~~|n-2m|=2\cos(\pi/\eta) \\
f(\theta) &=& f(i \pi-\theta);~~
f(\theta)f(-\theta)=[\sin(\frac{\pi-i\theta)}{\eta}
\sin(\frac{\pi +i\theta}{\eta})]^{-1}
\end{eqnarray}
and for solution (16) we have
\EQ
S_i(\theta)=f(\theta)
R_i(i\theta,\eta=\frac{2\pi}{n-2m-2}),~~f(\theta)f(-\theta)=\frac{\theta^2}
{\theta^2 +\eta^2}
\EN

The main feature of these $S$-matrices is that they have a formally remarkable
resemblance to those
describing the physics of $O(N)$ invariant systems\cite{AB,AB1}. Indeed,  at
$m=0$
the $O(n)$ symmetry is automatically
restored in the solutions (18-20). The physical
interpretation of these solutions is as follows. The
first solution can be considered as a regularized version of that
proposed by A.B. Zamolodchikov \cite{AB} to describe
the physics of a self avoiding polymer. In our case, however, we can choose
$n,m \neq 0$ such that the self avoiding limit $\eta=0$ is taken in an
unambiguous way. The second
solution (20) generalizes the $S$-matrices corresponding to
the $O(N)$ non-linear sigma model \cite{AB1}. An important feature is that now
the simplest case of $Osp(1|2)$ has pole on
the physical strip  at $\theta=i2\pi/3$(in function $f(\theta)$) which is not
present
in its equivalent  bosonic version, namely the $O(3)$ non-linear
sigma case. We believe that this is a very interesting solution and we hope to
discuss
its other features, e.g., the associated quantum spin chain and the quantum
field theory,
in a further publication.

\section*{Acknowledgements}
It is a pleasure to thank F.C. Alcaraz for discussions and a help with
numerical checks. This work is
supported by CNPq and Capes (Brazilian agencies).
\vspace{1.0cm}\\

\end{document}